\documentclass[conference]{IEEEtran}
\usepackage{graphicx}
\usepackage{amsmath,bbm,amssymb,amsfonts,amstext,amsopn,sansmath}
\usepackage{cite}
\usepackage{balance}
\usepackage{url}
\usepackage{epsfig}
\usepackage{setspace}
\usepackage{stmaryrd}
\usepackage{psfrag}	
\usepackage{multirow}
\usepackage{float}
\usepackage[process=auto]{pstool}
\usepackage{etoolbox}
\usepackage{algorithm}
\usepackage{algorithmic}
\allowdisplaybreaks
\usepackage[nolist]{acronym}

\newcommand{\x}{\mathsf{x}}
\newcommand{\y}{\mathsf{y}}
\newcommand{\z}{\mathsf{z}}

\newtheorem{remk}{Remark}

\newtheorem{prop}{Proposition}
\newtheorem{corol}{Corollary}

\newtoggle{OneColumn}
\toggletrue{OneColumn}

\newtoggle{arXiv}
\togglefalse{arXiv}

\EndPreamble
\begin{document}
\title{Physics-based Modeling  of Large Intelligent Reflecting Surfaces for Scalable Optimization \vspace{-0.3cm}}

\author{Marzieh Najafi$^\dag$,  Vahid Jamali$^\dag$, Robert Schober$^\dag$, and Vincent H. Poor$^\ddag$ \\
	$^\dag$Friedrich-Alexander University Erlangen-N\"urnberg, Germany\quad
	$^\ddag$Princeton University, USA \vspace{-0.4cm}
}

\maketitle

\vspace{-1.5cm}

\begin{abstract}
In this paper, we develop a physics-based model that allows a scalable optimization of large intelligent reflecting surfaces (IRSs). The basic idea is to partition the IRS unit cells into several subsets, referred to as tiles, and model the impact of each tile on the wireless channel. Borrowing concepts from the radar literature, we model each tile as an \textit{anomalous} reflector, and derive its impact on the wireless channel for given unit cell phase shifts by solving the corresponding integral equations for the electric and magnetic vector fields. Based on this model, one can design the phase shifts of the unit cells of a tile offline for the support of several transmission modes and then select the best mode online for a given channel realization. Therefore, the number of tiles and transmission modes in the proposed model are design parameters that can be adjusted to trade performance for complexity.
\end{abstract}

\acresetall
\section{Introduction}

Smart wireless environments are a newly emerging concept in wireless communications where intelligent reflecting surfaces (IRSs) are deployed to influence the propagation characteristics of the wireless channel \cite{di2019smart,liaskos2019novel}. IRSs consist of a large number of programmable sub-wavelength elements, so-called unit cells or meta atoms, that can change the properties of an impinging electromagnetic (EM) wave while reflecting it. For instance,  a properly designed unit cell phase distribution across the surface enables the IRS to alter the direction of the wavefront of the reflected wave, thereby realizing the generalized Snell's law \cite{estakhri2016wave,asadchy2016perfect}. 

As the physics-based models in \cite{ozdogan2019intelligent,bjornson2020power,di2020analytical} suggest, the path-loss of the end-to-end IRS-assisted links is significant for \textit{far-field} scenarios and indeed  a \textit{very large} IRS is needed to overcome it in practice. To see this, let $\rho_d$, $\rho_t$, and $\rho_r$ denote  the transmitter-to-receiver, transmitter-to-IRS, and IRS-to-receiver distances, respectively. Thereby, assuming free-space propagation, the path-losses of the direct and IRS-assisted links are proportional to $\frac{1}{4\pi\rho_d^2}$ and $\frac{A^2}{4\pi\lambda^2\rho_t^2\rho_r^2}$, respectively, where $\lambda$ is the wavelength and $A$ is the area of the IRS, see \cite{ozdogan2019intelligent} and Corollary~\ref{Corol:MaxContinuous} for details. Therefore, for the direct and IRS-assisted links to have similar path-losses, the normalized size of the IRS has to scale as $\frac{A}{\lambda^2}\propto  \frac{\rho_t\rho_r}{\rho_d\lambda}$. For example, assuming $\rho_d=200$~m, $\rho_t=\rho_r=100$~m, and a unit-cell spacing of half a wavelength, $Q\propto\frac{200}{\lambda}\approx 3300$ unit cells are needed for a carrier frequency of $5$~GHz. Therefore, the direct optimization of the unit cell phase shifts may not be a feasible approach for the online design of large~IRSs. 

The physics-based model derived in this paper generalizes the models in \cite{ozdogan2019intelligent,bjornson2020power,di2020analytical} which provide interesting insights, but were obtained under more restrictive assumptions. For instance, in  \cite{ozdogan2019intelligent}, the scatter field was characterized for a specific polarization and the angles of the impinging and reflected waves were in the same plane, see Remark~\ref{Remk:Special}. However, in practice, several waves may impinge on the same IRS from different directions and with different polarizations and will be redirected in different directions.  In \cite{bjornson2020power}, the authors studied the power scaling laws for asymptotically large IRSs; however, similar to \cite{ozdogan2019intelligent}, general incident and reflection directions were not considered. Furthermore, in \cite{di2020analytical}, the authors modeled an IRS in a \textit{two-dimensional} system using the \textit{scalar} theory of diffraction and the Huygens-Fresnel principle. In contrast, we consider a three-dimensional system and characterize the reflected vector field for all observation angles when a plane wave with arbitrary incident angle and arbitrary polarization impinges on the IRS.

In this paper, we develop a physics-based end-to-end channel model for \textit{large IRSs}  that accounts for all relevant effects and allows for the \textit{scalable optimization} of the IRS configuration.  The basic idea is to partition the $Q$ IRS unit cells into $N\ll Q$ tiles. We then model the impact of a tile on the wireless channel for a given phase-shift configuration of the unit cells of the tile.  In particular, borrowing an analogy from the radar literature, we model each tile as an \textit{anomalous} reflector, and assuming a far-field scenario, derive its response function by solving the corresponding integral equations for the electric and magnetic vector fields \cite{balanis2012advanced}.  Furthermore, we model the  IRS-assisted end-to-end channel between multiple transmitters and multiple receivers as a function of the response functions of all tiles of the IRS where each tile is set to support a given transmission mode. The number of parameters for this model scales with the number of tiles $N$ and the number of transmission modes that each tile supports, denoted by $M$. Therefore, the search space for online optimization of the IRS-assisted system does not scale with $Q$ but with  $N$ and  $M$, which are design parameters and can be adjusted to trade performance for complexity. 

\section{IRS Structure}\label{Sec:IRSstruct}

\begin{figure}[t]\vspace{-0.5cm}
	\centering
	\includegraphics[width=0.5\textwidth]{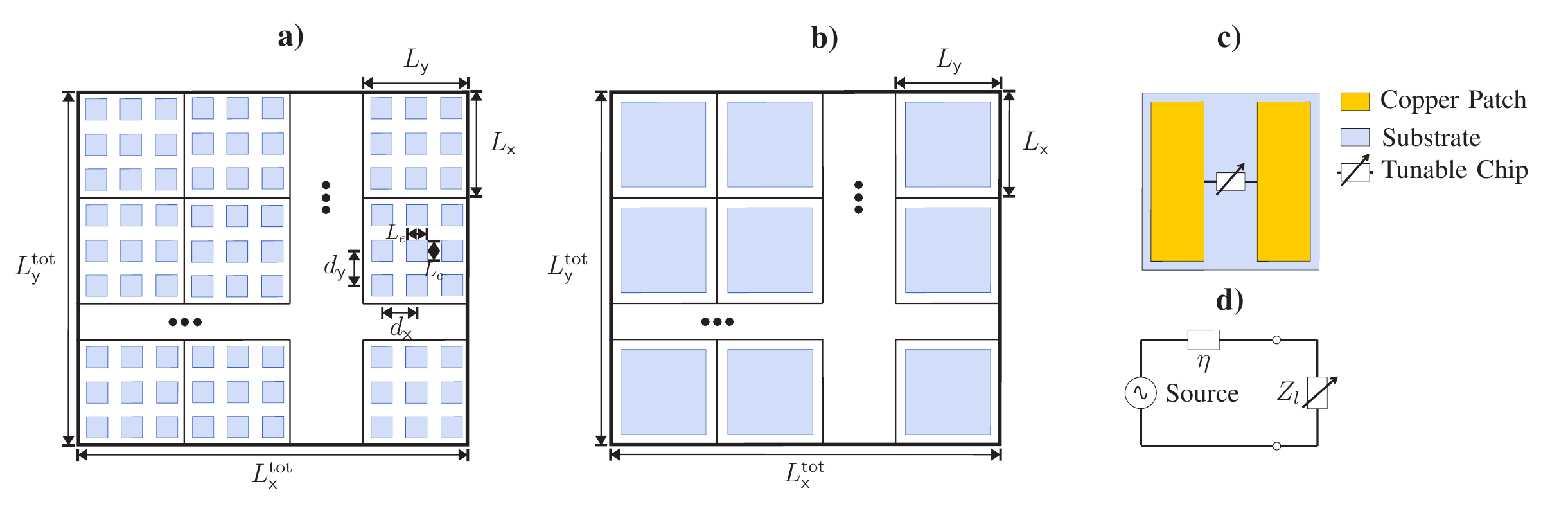}\vspace{-0.3cm}
	\caption{Schematic illustration of planar IRS of size $L_{\x}^{\mathrm{tot}}\times L_{\y}^{\mathrm{tot}}$ partitioned into tiles of size $L_\x \times L_\y$. a) Each tile is composed of square unit cells of size $L_e\times L_e$ which are spaced by $d_\x$ and $d_\y$ along the $x$ and $y$ directions, respectively. b) Each tile may be approximated as a continuous surface. c) Schematic illustration of tunable unit cells. d) Equivalent circuit model for the unit cells, see \cite{asadchy2016perfect,abeywickrama2019intelligent} for implementations of the tunable chip.\vspace{-0.2cm}} 
	\label{Fig:IRS}
\end{figure}

We consider a large rectangular IRS of size $L_{\x}^{\mathrm{tot}}\times L_{\y}^{\mathrm{tot}}$ placed in the $x-y$ plane where $L_\x^{\mathrm{tot}},L_\y^{\mathrm{tot}}\gg \lambda$, see Fig.~\ref{Fig:IRS}.  The IRS is composed of many sub-wavelength unit cells (also known as meta atoms) of size $L_e\times L_e$ that are able to change the properties of an impinging EM wave when reflecting it. Typically, each unit cell contains programmable components  (such as tunable varactor diodes or switchable positive-intrinsic-negative (PIN) diodes) that can change the reflection coefficient of the surface, which we denote by $\Gamma$,  see Fig.~\ref{Fig:IRS} c). 


We assume that the IRS is partitioned into tiles of size $L_\x \times L_\y$. For notational simplicity, let us assume that $L_{\x}^{\mathrm{tot}}/L_{\x}$ and $L_{\y}^{\mathrm{tot}}/L_{\y}$ are integers and in total, there are $N=L_{\x}^{\mathrm{tot}}L_{\y}^{\mathrm{tot}}/(L_{\x}L_{\y})$ tiles. Each tile consists of several programmable sub-wavelength unit cells. Here, assuming a unit-cell spacing of $d_\x$ and $d_\y$ along the $x$ and $y$ axes, respectively, the total number of unit cells of the IRS is given by $Q=NQ_\x Q_\y$, where $Q_\x=L_\x/d_\x$ and $Q_\y=L_\y/d_\y$. When $d_\x=d_\y\approx L_e\ll \lambda$ and $L_\x,L_\y\gg \lambda$, the collection of all unit cells on one tile acts as a continuous programmable surface \cite{estakhri2016wave,asadchy2016perfect}, cf. Fig.~\ref{Fig:IRS} b). In this paper, an ideal tile that acts as a continuous programmable surface is referred to as a \textit{continuous tile}. In contrast, a practical tile that comprises a finite number of unit cells is referred to as a \textit{discrete tile}. We use the notion of continuous tiles in Section III to rigorously analyze the reflected EM field. 

\begin{figure}[t] 
	\begin{minipage}[c]{0.59\linewidth}
		\centering
		\includegraphics[width=0.9\linewidth]{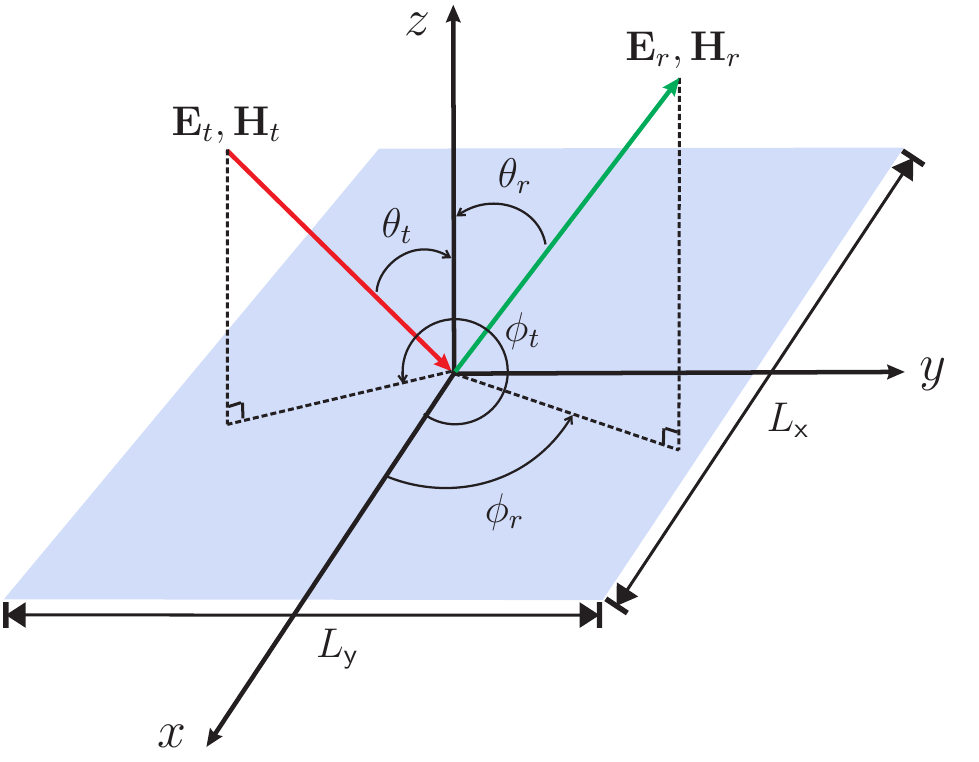}\vspace{-0.3cm}
	\end{minipage}
	\begin{minipage}[c]{0.4\linewidth}\vspace{-0.6cm} 
		\caption{Uniform plane wave impinges on a rectangular conducting tile with incident angle $\boldsymbol{\Psi}_t=(\theta_t,\phi_t,\varphi_t)$ and is reflected with a desired reflection angle $\boldsymbol{\Psi}_r=(\theta_r,\phi_r)$. }
		\label{Fig:CoordinatesContinuous}
	\end{minipage}\vspace{-0.2cm}
\end{figure}

\section{Tile Response Function}
We focus on the far-field scenario where the curvature of the wavefront originating from a distant source on a tile can be neglected. Therefore, the incident EM wave can be modeled as a plane wave impinging on the tile and is characterized by $\boldsymbol{\Psi}_t=(\theta_t,\phi_t,\varphi_t)$, see Fig.~\ref{Fig:CoordinatesContinuous}. Here, $\theta_t$ and $\phi_t$ denote the elevation and azimuth angles of the incident direction, respectively, and $\varphi_t$ determines the polarization of the incident wave. The unit cells of the tile act as secondary sources and reflect the incident EM wave. The signal received at a receiver in the far field of the tile can be characterized by the overall complex tile response function $g(\boldsymbol{\Psi}_t,\boldsymbol{\Psi}_r)$, where  $\boldsymbol{\Psi}_r=(\theta_r,\phi_r)$ denotes the reflection angle  at which the receiver is located.
The power of the tile response function, $|g(\boldsymbol{\Psi}_t,\boldsymbol{\Psi}_r)|^2$, is  referred to as the radar cross section (RCS) of an object \cite[p.~584]{balanis2012advanced}. Here, to be able to model the superposition of multiple waves at a receiver in the far field, we generalize the concept of RCS to also include the phase information, i.e., 
\begin{IEEEeqnarray}{rll}\label{Eq:Gdef}
	g(\boldsymbol{\Psi}_t,\boldsymbol{\Psi}_r)	= \underset{\rho_r\to\infty}{\mathrm{lim}} \,\, \sqrt{4\pi \rho_r^2} e^{-jk\rho_r} \frac{E_r(\boldsymbol{\Psi}_r)}{E_t(\boldsymbol{\Psi}_t)},
\end{IEEEeqnarray}
where $k=\frac{2\pi}{\lambda}$ is the wave number, $E_t(\boldsymbol{\Psi}_t)$ is a phasor denoting the complex amplitude of the incident electric field impinging from angle $\boldsymbol{\Psi}_t$ on the tile center (i.e., $(x,y)=(0,0)$),  and $E_r(\boldsymbol{\Psi}_r)$ is a phasor denoting the complex amplitude of  the reflected electric field in direction $\boldsymbol{\Psi}_r$ and at distance $\rho_r$ from the tile center.  

In order to study the impact of a continuous tile on an impinging EM wave, we first explicitly define the incident electric and magnetic fields. Here, we assume the following incident fields with arbitrary polarization and incident angle \cite[Ch.~11]{balanis2012advanced}\footnote{In \cite{balanis2012advanced},  the incident wave is always assumed to be in the $y-z$ plane and the polarization is either transverse electric $x$ (TE$^x$) or transverse magnetic $x$ (TM$^x$) to facilitate the analysis. While these assumptions are without loss of generality when analyzing one impinging wave, in this paper, we deal with scenarios where multiple waves may arrive from different angles and with different polarizations, and hence, these simplifying assumptions cannot simultaneously hold for all impinging waves. Therefore, we generalize the formulation of the electric and magnetic fields in \cite{balanis2012advanced} to arbitrary incident angles and polarizations.}   
\begin{IEEEeqnarray}{lll}\label{Eq:IncidentFields}
	\mathbf{E}_t(\boldsymbol{\Psi}_t) = E_0  e^{j k \mathbf{a}_t \cdot (\mathbf{e}_\x x+ \mathbf{e}_\y y + \mathbf{e}_\z z)} \mathbf{a}_E
	\IEEEyesnumber\IEEEyessubnumber\\
	\mathbf{H}_t(\boldsymbol{\Psi}_t) =  \frac{E_0}{\eta}  e^{jk \mathbf{a}_t \cdot (\mathbf{e}_\x x+ \mathbf{e}_\y y + \mathbf{e}_\z z)} \mathbf{a}_H,\IEEEyessubnumber
\end{IEEEeqnarray}
where $E_0$ is the magnitude of the incident electric field, $\mathbf{a}\cdot\mathbf{b}$ denotes the inner product of vectors $\mathbf{a}$ and $\mathbf{b}$, $\eta=\sqrt{\frac{\mu}{\epsilon}}$ is the characteristic impedance, $\mu$ is the magnetic permeability, and $\epsilon$ is the electric permittivity. Moreover, $\mathbf{e}_\x$, $\mathbf{e}_\y$, and $\mathbf{e}_\z$ denote the unit vectors in the $x$, $y$, and $z$ directions, respectively, and $\mathbf{a}_E$, $\mathbf{a}_H$, and $\mathbf{a}_t$ denote the directions of the electric field, the magnetic field, and the propagation of the incident wave, respectively, which are all mutually orthogonal. 
In spherical coordinates, the incident direction is defined as
\begin{IEEEeqnarray}{lll}\label{Eq:Incident}
	\mathbf{a}_t
	&=(\sin(\theta_t)\cos(\phi_t),\sin(\theta_t)\sin(\phi_t),\cos(\theta_t))\nonumber\\ &\triangleq(A_\x(\boldsymbol{\Psi}_t),A_\y(\boldsymbol{\Psi}_t),A_\z(\boldsymbol{\Psi}_t)).
\end{IEEEeqnarray}
Note that $\mathbf{a}_E$ and $\mathbf{a}_H$ lie in the plane orthogonal to $\mathbf{a}_t$. Let $(H_\x,H_\y)$ denote the components of the magnetic field in the $x-y$ plane. Defining $\varphi_t=\tan^{-1}(\frac{H_\y}{H_\x})$, which determines  the polarization of the incident wave, we obtain $\mathbf{a}_E$ and $\mathbf{a}_H$  as
\begin{IEEEeqnarray}{lll} 
	\mathbf{a}_E= \mathbf{a}_H\times \mathbf{a}_t \quad \text{and} 	\IEEEyesnumber\IEEEyessubnumber \\ \mathbf{a}_H = b  \big(c(\boldsymbol{\Psi}_t)\cos(\varphi_t),c(\boldsymbol{\Psi}_t)\sin(\varphi_t),\sqrt{1-c^2(\boldsymbol{\Psi}_t)}\big), \quad	 \IEEEyessubnumber
\end{IEEEeqnarray}
where  $c(\boldsymbol{\Psi}_t)=\frac{A_\z(\boldsymbol{\Psi}_t)}{\sqrt{A_{\x,\y}^2(\boldsymbol{\Psi}_t)+A_\z^2(\boldsymbol{\Psi}_t)}}$,  $A_{\x,\y}(\boldsymbol{\Psi}_t) = \cos(\varphi_t) A_\x(\boldsymbol{\Psi}_t) + \sin(\varphi_t) A_\y(\boldsymbol{\Psi}_t)$,  $b=\mathrm{sign}\big(\frac{H_\x}{c(\boldsymbol{\Psi}_t)\cos(\varphi_t)}\big)$, $\mathrm{sign}(\cdot)$ denotes the sign of a real number, and $\mathbf{a}\times\mathbf{b}$ denotes the cross product between vectors $\mathbf{a}$ and $\mathbf{b}$. Note that the reference complex amplitude of the incident electric field in \eqref{Eq:Gdef} can be obtained from the electric vector field in  \eqref{Eq:IncidentFields} as $E_t(\boldsymbol{\Psi}_t)=E_0  e^{j k \mathbf{a}_t \cdot (\mathbf{e}_\x x+ \mathbf{e}_\y y + \mathbf{e}_\z z)} \big|_{(x,y,z)=(0,0,0)}=E_0$.

We assume that the surface impedance is suitably designed to realize reflection coefficient $\Gamma=\rho_{\rm eff}e^{j\beta(x,y)}$, where $\beta(x,y)$ is the phase shift applied at point $(x,y)$ on the tile and $\rho_{\rm eff}\in(0,1]$ is the efficiency factor of the tile which accounts for potential power losses \cite{estakhri2016wave,asadchy2016perfect}.
Therefore, the tangential components of the scattered electric and magnetic fields are given by $\mathbf{E}_r=\Gamma\mathbf{E}_t$ and $\mathbf{H}_r=-\Gamma\mathbf{H}_t$, respectively. In order to determine the scattered fields, we employ the Electromagnetic Equivalence Theorem \cite[Ch.~7]{balanis2012advanced} and assume that \textit{only} scattered fields $(\mathbf{E}_r,\mathbf{H}_r)$  exist in the  environment and that the IRS is replaced by a perfectly magnetically conducting (PMC) surface. To compensate for the field discontinuity across the boundaries, an electric current $\mathbf{J}_r^{\mathrm{pmc}} = \mathbf{n}\times\mathbf{H}_r\big|_{z=0}$ should be introduced on the surface, where $\mathbf{n}$ is the normal vector of the surface \cite[Ch.~7, eq. (7.42)]{balanis2012advanced}. Next, using the Image Theory for large flat surfaces \cite[Ch.~7.4]{balanis2012advanced}, an equivalent \textit{obstacle-free}  system is obtained by removing the PMC and replacing $\mathbf{J}_r^{\mathrm{pmc}}$ with an equivalent electric current $\mathbf{J}_r$  obtained as  \cite[Ch.~7.4, 7.8]{balanis2012advanced}. 
\begin{IEEEeqnarray}{lll}\label{Eq:Current_t}
	\mathbf{J}_r &=2\mathbf{J}_r^{\mathrm{pmc}} = 2\mathbf{n}\times\mathbf{H}_r\big|_{z=0}
	= -2\Gamma\mathbf{n}\times\mathbf{H}_t\big|_{z=0}\nonumber\\
	&\overset{(a)}{=}  -2\rho_{\rm eff}\frac{E_0}{\eta} e^{jk[A_\x(\boldsymbol{\Psi}_t)x+A_\y(\boldsymbol{\Psi}_t)y]+j\beta(x,y)} \mathbf{e}_\z \times \mathbf{a}_H  \nonumber\\
	&\overset{(b)}{=}    e^{jk[A_\x(\boldsymbol{\Psi}_t)x+A_\y(\boldsymbol{\Psi}_t)y]+j\beta(x,y)} (J_\x\mathbf{e}_\x + J_\y\mathbf{e}_\y),
\end{IEEEeqnarray}
where equality $(a)$ follows from the assumption that the incident EM wave is a plane wave, and for equality $(b)$, we used the definitions
$J_\x = 2 \frac{E_0}{\eta}\rho_{\rm eff} c(\boldsymbol{\Psi}_t)\sin(\varphi_t)$ and $J_\y = -2 \frac{E_0}{\eta}\rho_{\rm eff} c(\boldsymbol{\Psi}_t)\cos(\varphi_t)$. The magnitude of the equivalent electric current is given by $\|\mathbf{J}_r\| = \sqrt{J_\x^2+J_\y^2} = 2 \rho_{\rm eff}\frac{E_0}{\eta} c(\boldsymbol{\Psi}_t)$, where depending on the incident angle and the polarization, we have $c(\boldsymbol{\Psi}_t)\in[\cos(\theta_t),1]$.

The reflected electric and magnetic fields induced by electric current $\mathbf{J}_r$ in an obstacle-free environment are found as follows \cite[Ch.~6]{balanis2012advanced} 
\begin{IEEEeqnarray}{lll}\label{Eq:HEV}
	\mathbf{E}_r = \frac{1}{j\omega\epsilon}\nabla \times \mathbf{H}_r\quad\text{and}\quad \mathbf{H}_r  = \frac{1}{\mu} \nabla \times \mathbf{V}, 
\end{IEEEeqnarray}
where $\nabla \times$ is the curl operator, $\omega=k/\sqrt{\mu\epsilon}$, and $\mathbf{V}$ is an auxiliary vector potential, which assuming a far-field scenario is given by 
\begin{IEEEeqnarray}{lll}\label{Eq:EintegralFF}
	\mathbf{V}(\boldsymbol{\Psi}_r) \!=\!  \frac{\mu e^{-jk\rho_r}}{4\pi \rho_r} \!\!\int_{-\frac{L_\x}{2}}^{\frac{L_\x}{2}}\!\int_{-\frac{L_\y}{2}}^{\frac{L_\y}{2}} \!\!\mathbf{J}_r(x,y) e^{jk \sqrt{x^2+y^2} \cos(\alpha)} \mathrm{d}x\mathrm{d}y.\quad\,
\end{IEEEeqnarray}
Here, $\alpha$ is the angle between the vector specified by angle $\boldsymbol{\Psi}_r$ and the line that connects $(x,y)$ with the origin. 

In order to solve the integral equation in \eqref{Eq:EintegralFF}, we have to assume a given phase-shift profile, $\beta(x,y)$, for the tile surface, i.e., a transmission mode. One criterion to design a tile transmission mode is to facilitate reflection in a certain direction, i.e., the generalized Snell's law~\cite{kaina2014shaping,najafi2019intelligent}. In particular, we design the tile to reflect an EM wave impinging from direction $\boldsymbol{\Psi}_t^*$ towards direction $\boldsymbol{\Psi}_r^*$ and analyze the tile response function $g(\boldsymbol{\Psi}_t,\boldsymbol{\Psi}_r)$ caused by the corresponding phase-shift profile for an EM wave impinging from an arbitrary direction $\boldsymbol{\Psi}_t$ (including $\boldsymbol{\Psi}_t^*$) and observed at an arbitrary observation angle $\boldsymbol{\Psi}_r$ (including $\boldsymbol{\Psi}_r^*$). For ease of presentation, let us define the amplitude and phase of $g(\boldsymbol{\Psi}_t,\boldsymbol{\Psi}_r)$ as $g_{||}(\boldsymbol{\Psi}_t,\boldsymbol{\Psi}_r)$ and $g_{\angle}(\boldsymbol{\Psi}_t,\boldsymbol{\Psi}_r)$, respectively, up to a sign, i.e., $g_{||}(\boldsymbol{\Psi}_t,\boldsymbol{\Psi}_r)=\pm |g(\boldsymbol{\Psi}_t,\boldsymbol{\Psi}_r)|$ and $g_{\angle}(\boldsymbol{\Psi}_t,\boldsymbol{\Psi}_r)=\angle g(\boldsymbol{\Psi}_t,\boldsymbol{\Psi}_r)\pm \pi$ such that $g(\boldsymbol{\Psi}_t,\boldsymbol{\Psi}_r)=g_{||}(\boldsymbol{\Psi}_t,\boldsymbol{\Psi}_r)e^{j g_{\angle}(\boldsymbol{\Psi}_t,\boldsymbol{\Psi}_r)}$. Here, $|\cdot|$ and $\angle$ denote the absolute value and phase of a complex number. Moreover, let $A_i(\boldsymbol{\Psi}_t,\boldsymbol{\Psi}_r)=A_i(\boldsymbol{\Psi}_t)+A_i(\boldsymbol{\Psi}_r),i\in\{\x,\y\}$. 

\begin{figure*}[t]\vspace{-0.5cm}
	\begin{minipage}[c]{0.65\linewidth}
		\hspace{-1cm}
		\includegraphics[width=1.1\linewidth]{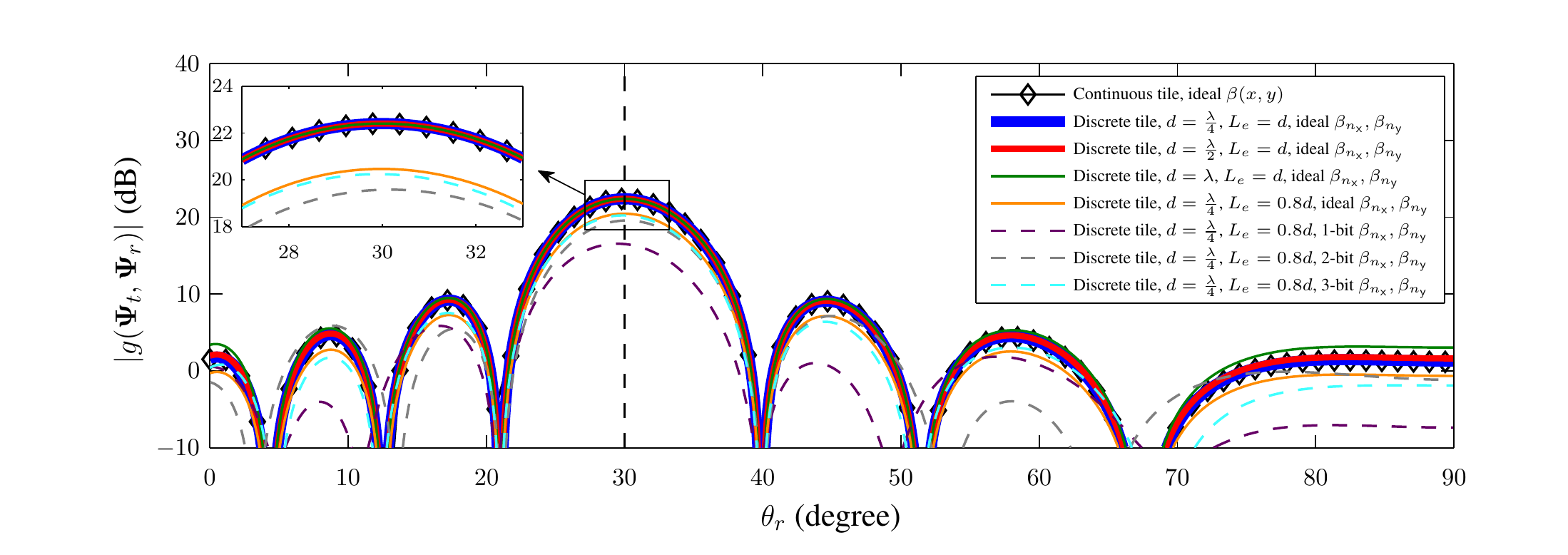}\vspace{-0.3cm}
	\end{minipage}
	\begin{minipage}[c]{0.35\linewidth}\vspace{-0.3cm} 
		\caption{Amplitude of the tile response function, $|g(\boldsymbol{\Psi}_t,\boldsymbol{\Psi}_r)|$, in dB vs. $\theta_r$  for $(\theta_t,\phi_t,\varphi_t)=(0,0,22.5^{\circ})$, $\phi_r=45^{\circ}$, $(\theta_t^*,\phi_t^*)=(0,0)$, $(\theta_r^*,\phi_r^*)=(30^{\circ},45^{\circ})$, $d_\x=d_\y=d$, and $\rho_{\rm eff}=0.5$. For a fair comparison, the sizes of the continuous and discrete tiles are identical with $L_\x=L_\y=10\lambda$. For discrete tiles, the number of unit cells along the $\x$ and $\y$ axes are $Q_\x=\frac{L_\x}{d_\x}$ and $Q_\y=\frac{L_\y}{d_\y}$, respectively, the unit cell size is $L_e\times L_e$, and we consider both ideal continuous and uniformly quantized phase shifts $\beta_{n_\x}$ and $\beta_{n_\y}$.\vspace{-0.5cm} }
		\label{Fig:FuncGdiscrete}
	\end{minipage}
\end{figure*}

\begin{prop}\label{Prop:Continuous}
	For given $\boldsymbol{\Psi}_t^*$ and $\boldsymbol{\Psi}_r^*$, let us impose linear phase-shift function $\beta(x,y)=\beta(x)+\beta(y)$ with
	\begin{IEEEeqnarray}{rll}\label{Eq:BetaContinuous}
		\beta(x) \!=\! -kA_\x(\boldsymbol{\Psi}_t^*,\boldsymbol{\Psi}_r^*)x \!+\!\frac{\beta_0}{2},
		\beta(y) \!=\! -kA_\y(\boldsymbol{\Psi}_t^*,\boldsymbol{\Psi}_r^*)y \!+\! \frac{\beta_0}{2}\quad\,
	\end{IEEEeqnarray}
	Then, the amplitude of the corresponding tile response function for an EM wave impinging from an arbitrary direction $\boldsymbol{\Psi}_t$  and observed at arbitrary reflection direction $\boldsymbol{\Psi}_r$ is obtained~as  
	\begin{IEEEeqnarray}{lll}\label{Eq:AmplContinuous}
		g_{||}(\boldsymbol{\Psi}_t,\boldsymbol{\Psi}_r) = \frac{\sqrt{4\pi} \rho_{\rm eff} L_\x L_\y}{\lambda}\widetilde{g}(\boldsymbol{\Psi}_t,\boldsymbol{\Psi}_r) \nonumber \\
		\quad \times \mathrm{sinc}\left(\frac{k L_\x [A_\x(\boldsymbol{\Psi}_t,\boldsymbol{\Psi}_r) - A_\x(\boldsymbol{\Psi}_t^*,\boldsymbol{\Psi}_r^*)]}{2}\right)\nonumber \\
		\quad\times \mathrm{sinc}\left(\frac{k L_\y [A_\y(\boldsymbol{\Psi}_t,\boldsymbol{\Psi}_r) - A_\y(\boldsymbol{\Psi}_t^*,\boldsymbol{\Psi}_r^*)]}{2}\right),\quad 
	\end{IEEEeqnarray}
	where $\mathrm{sinc}(x)=\sin(x)/x$ and
	\begin{IEEEeqnarray}{lll} \label{Eq:Gtilde}
		\widetilde{g}(\boldsymbol{\Psi}_t,\boldsymbol{\Psi}_r) = c(\boldsymbol{\Psi}_t) \nonumber \\
		\times \left\|
		\begin{bmatrix}
			\cos(\varphi_t)\cos(\theta_r)\sin(\phi_r)-\sin(\varphi_t)\cos(\theta_r)\cos(\phi_r)\\
			\sin(\varphi_t)\sin(\phi_r)+\cos(\varphi_t)\cos(\phi_r)
		\end{bmatrix}
		\right\|_2\quad
	\end{IEEEeqnarray}
	The phase of the tile response function is obtained as  $g_{\angle}(\boldsymbol{\Psi}_t,\boldsymbol{\Psi}_r) = \frac{\pi}{2}+\beta_0$. 
\end{prop}
\begin{IEEEproof}
The proof follows similar steps as that provided in \cite[p. 591-597]{balanis2012advanced} and is omitted here due to space constraints. Interested readers are referred to \cite[Appendix~A]{najafi2020intelligent} which is the journal version of this paper and contains the proof.
\end{IEEEproof}

\begin{corol}\label{Corol:MaxContinuous}
	Assuming large $L_\x,L_\y\gg \lambda$ and $\boldsymbol{\Psi}_t^*=\boldsymbol{\Psi}_t$, the maximum value of $|g(\boldsymbol{\Psi}_t,\boldsymbol{\Psi}_r)|$ is observed at $\boldsymbol{\Psi}_r=\boldsymbol{\Psi}_r^*$ and is given by
	\begin{IEEEeqnarray}{rll}\label{Eq:max_cont}
		|g(\boldsymbol{\Psi}_t,\boldsymbol{\Psi}_r)| 
		\!=\! \frac{\sqrt{4\pi} \rho_{\rm eff} L_\x L_\y}{\lambda}\widetilde{g}(\boldsymbol{\Psi}_t^*,\boldsymbol{\Psi}_r^*) 
		\!\overset{(a)}{\leq}\! \frac{\sqrt{4\pi} \rho_{\rm eff} L_\x L_\y}{\lambda},\quad\,\,\, 
	\end{IEEEeqnarray}
	where $(a)$ holds with equality for incident and reflection directions normal to the surface, i.e., $\theta_t=\theta_r=0$.
\end{corol}
\begin{IEEEproof}
	The proof follows from computing the limits of \eqref{Eq:AmplContinuous} as $\boldsymbol{\Psi}_r\to\boldsymbol{\Psi}_r^*$ and simplifying the result for normal incident and reflection directions.
\end{IEEEproof}

\begin{remk}\label{Remk:Special}
	The result given in  \cite[Lemma~2]{ozdogan2019intelligent} is a special case of Proposition~\ref{Prop:Continuous} with $\phi_t=\frac{3\pi}{2}$, $\varphi_t=\frac{\pi}{2}$, $\phi_r^*=\phi_r=\frac{\pi}{2}$,  $\rho_{\mathrm{eff}}=1$, and $A_i(\boldsymbol{\Psi}_t^*,\boldsymbol{\Psi}_r^*)=\beta_0=0,\,\,i\in\{\x,\y\}$, which implies  $\widetilde{g}(\boldsymbol{\Psi}_t,\boldsymbol{\Psi}_r) = c(\boldsymbol{\Psi}_t)=\cos(\theta_t)$,  $A_\x(\boldsymbol{\Psi}_t,\boldsymbol{\Psi}_r)=0$ and $A_\y(\boldsymbol{\Psi}_t,\boldsymbol{\Psi}_r)=\sin(\theta_r)-\sin(\theta_t)$.
\end{remk}

 \begin{remk}\label{Remk:Discrete}
 	 The response function $g(\boldsymbol{\Psi}_t,\boldsymbol{\Psi}_r)$ for  discrete tiles can be found as the superposition of the response functions of all individual unit cells, see \cite{tang2019wireless,najafi2020intelligent}, and \cite[Ch. 3]{lau2012reconfigurable} for similar models. Thereby, the discrete phase shift at the $(n_\x,n_\y)$-th unit cell is denoted by $\beta_{n_\x,n_\y}=\beta_{n_\x}+\beta_{n_\y}$.  Exploiting Proposition~\ref{Prop:Continuous}, assuming $L_e\ll\lambda$, and using the identity $\lim_{x\to 0} \mathrm{sinc}(x)=1$, we can characterize the response of an individual unit cell, denoted by $g_{n_\x,n_\y}(\boldsymbol{\Psi}_t,\boldsymbol{\Psi}_r),\,\,\forall n_\x,n_\y$,~as
 	\begin{IEEEeqnarray}{ll} 
 		g_{n_\x,n_\y}(\boldsymbol{\Psi}_t,\boldsymbol{\Psi}_r)	= 
 		\frac{j\sqrt{4\pi} \rho_{\rm eff} L_e^2}{\lambda} \, \widetilde{g}(\boldsymbol{\Psi}_t,\boldsymbol{\Psi}_r)
 		\nonumber\\
 		\times e^{jkd_\x [ A_\x(\boldsymbol{\Psi}_t)+ A_\x(\boldsymbol{\Psi}_r)]n_\x+j\beta_{n_\x}} 
 		\, e^{jkd_\y [A_\y(\boldsymbol{\Psi}_t)+ A_\y(\boldsymbol{\Psi}_r)]n_\y+j\beta_{n_\y}}.\quad\,\,\,\,
 	\end{IEEEeqnarray}
 	The tile response function  of the entire tile is the superposition of the response functions of all its unit cells \cite[Ch. 3]{lau2012reconfigurable} and is obtained as
 	\begin{IEEEeqnarray}{rll}\label{Eq:Discrete}
 		g(\boldsymbol{\Psi}_t,\boldsymbol{\Psi}_r)	=  \sum_{n_\x=-\frac{Q_\x}{2}+1}^{\frac{Q_\x}{2}}  \sum_{n_\y=-\frac{Q_\y}{2}+1}^{\frac{Q_\y}{2}}  g_{n_\x,n_\y}(\boldsymbol{\Psi}_t,\boldsymbol{\Psi}_r).
 	\end{IEEEeqnarray}
 \end{remk}

  Fig.~\ref{Fig:FuncGdiscrete} shows the amplitude of the tile response function for an anomalous reflection for both continuous and discrete tiles.  We now highlight some insights from Proposition~\ref{Prop:Continuous}, Corollary~\ref{Corol:MaxContinuous}, Remark~\ref{Remk:Discrete}, and Fig.~\ref{Fig:FuncGdiscrete}: 

\textit{i)} Eq.~\eqref{Eq:AmplContinuous}  show that $|g(\boldsymbol{\Psi}_t,\boldsymbol{\Psi}_r)|$ becomes narrower as $L_\x$ and $L_\y$ increase. However, even for large tiles of size $L_\x=L_\y=10\lambda$, the $10$-dB beamwidth\footnote{Here, the $10$-dB beamwidth is defined as the maximum range of $\theta_r$ around $\theta_r^*$ for which $|g(\boldsymbol{\Psi}_t,\boldsymbol{\Psi}_r)|$ is not more than $10$~dB smaller than its maximum value.} is around $15$ degree (see Fig.~\ref{Fig:FuncGdiscrete}) which can cause significant interference to unintended receivers in far-field scenarios. 

\textit{ii)} For the phase-shift function in \eqref{Eq:BetaContinuous}, the phase of $g(\boldsymbol{\Psi}_t,\boldsymbol{\Psi}_r)$ is equal to $\beta_0$ up to a constant. In other words, if we change the phase induced on the \textit{entire tile surface} by a constant,  $|g(\boldsymbol{\Psi}_t,\boldsymbol{\Psi}_r)|$ remains the same and $\angle g(\boldsymbol{\Psi}_t,\boldsymbol{\Psi}_r)$ changes by that constant. This result can be exploited for coherent superposition of the reflected fields  of different tiles. 


\textit{iii)} Fig.~\ref{Fig:FuncGdiscrete}  shows that for a discrete tile to accurately approximate a continuous tile,  it is sufficient that $L_e=d_\x=d_\y\leq \frac{\lambda}{2}$ holds.  In practice, there are gaps between the IRS unit cells, i.e., $L_e<d_\x,d_\y$. This leads to a decrease of the effective size of the tile, see Fig.~\ref{Fig:FuncGdiscrete} for $L_e=0.8d$. 

\textit{iv)} Till now, we have assumed that the phase shift introduced by the tile unit cells can assume any real value which is an idealized assumption as finite resolution phase shifts are typically applied in practice. Nevertheless, Fig.~\ref{Fig:FuncGdiscrete}  suggests that a 3-bit uniform quantization of the phase shifts $\beta_{n_\x}$ and $\beta_{n_\y}$ yields a tile response function which is very close to the one obtained for ideal real-valued phase shifts. Moreover, from Fig.~\ref{Fig:FuncGdiscrete}, we also observe that even for a 1-bit phase shift quantization, the tile response function has a shape similar to the ideal case although the peak is reduced and the side lobes  deviate from the ideal real-valued phase shifts, which is consistent with  \cite{kaina2014shaping}.

\section{End-to-End Channel Model}

Using the tile response function $g(\boldsymbol{\Psi}_t,\boldsymbol{\Psi}_r)$ derived in the previous section, we now develop an  end-to-end channel model for IRS-assisted wireless systems comprising multiple transmitters and multiple receivers. In particular, 
we consider a general system comprising $N_t$ multiple-antenna transmitters, an IRS, and $N_r$ multiple-antenna receivers. In addition, we assume there are multiple scatterers in the environment which cause the signal of a given transmitter to arrive at the IRS potentially via multiple paths and the signal reflected from the IRS to potentially also arrive at a given receiver via multiple paths \cite{jamali2018scalable}. The resulting end-to-end channel model can be compactly written as follows 
\begin{IEEEeqnarray}{lll}\label{Eq:E2Emodel}
\mathbf{y}^{(j)} \!= &\sum_{i=1}^{N_t} 
\Big[\underset{\text{direct paths}}{\underbrace{\mathbf{A}_d^{(j,i)}\boldsymbol{\Sigma}_d^{(j,i)} \mathbf{D}_d^{(j,i)\mathsf{H}}}}  \\
&+\underset{\text{IRS-guided paths}}{\underbrace{\mathbf{A}_r^{(j)} \boldsymbol{\Sigma}_r^{(j)} \mathbf{G}^{(j,i)} \boldsymbol{\Sigma}_t^{(i)} \mathbf{D}_t^{(i)\mathsf{H}}}}\Big] \mathbf{x}^{(i)}\!+\!\mathbf{z}^{(j)}, \, j=1,\dots,N_r,\qquad\hspace{-0.7cm}\nonumber
\end{IEEEeqnarray}
where  $(\cdot)^{\mathsf{H}}$ denotes the Hermitian operator, $\mathbf{x}^{(i)}\in\mathbb{C}^{T_i}$ is the transmit symbol vector of the $i$-th transmitter which is equipped with $T_i$ transmit antennas; $\mathbf{y}^{(j)}\in\mathbb{C}^{J_j}$ denotes the receive vector at the $j$-th receiver which is equipped with $J_j$ receive antennas; and  $\mathbf{z}^{(j)}\in\mathbb{C}^{J_j}$ denotes the additive white Gaussian noise (AWGN) at the $j$-th receiver. Here, $\mathbb{C}$ denotes the set of complex numbers.  
Moreover, 
$\mathbf{A}_r^{(j)}\in\mathbb{C}^{J_j\times L_r^{(j)}}$ and $\mathbf{D}_t^{(i)}\in\mathbb{C}^{T_i\times L_t^{(i)}}$   ($\mathbf{A}_d^{(j,i)}\in\mathbb{C}^{J_j\times L_d^{(j,i)}}$ and $\mathbf{D}_d^{(j,i)}\in\mathbb{C}^{T_i\times L_d^{(j,i)}}$) are  matrices whose columns are the receive and transmit steering vectors evaluated at the angles-of-arrival (AoAs) and angles-of-departure (AoDs) of the IRS-assisted paths (direct paths),  respectively, where $L_d^{(j,i)}$, $L_t^{(i)}$, and $L_r^{(j)}$  denote the numbers 
of scatterers for the transmitter~$i$-to-receiver~$j$,
transmitter~$i$-to-IRS, and IRS-to-receiver~$j$ links, respectively. 
Furthermore,
$\boldsymbol{\Sigma}_d^{(j,i)}\in\mathbb{C}^{L_d^{(j,i)}\times L_d^{(j,i)}}$, $\boldsymbol{\Sigma}_t^{(i)}\in\mathbb{C}^{L_t^{(i)}\times L_t^{(i)}}$, and $\boldsymbol{\Sigma}_r^{(j)}\in\mathbb{C}^{L_r^{(j)}\times L_r^{(j)}}$ are diagonal matrices containing the channel coefficients of the transmitter~$i$-to-receiver~$j$, transmitter~$i$-to-IRS, and IRS-to-receiver $j$ paths, respectively. Note that $\boldsymbol{\Sigma}_d^{(j,i)}$, $\boldsymbol{\Sigma}_t^{(i)}$, and $\boldsymbol{\Sigma}_r^{(j)}$ contain the impact of path-loss as well as small-scale and large-scale fading. Finally, matrix $\mathbf{G}^{(j,i)}\in\mathbb{C}^{L_r^{(j)}\times L_t^{(i)}}$ contains the tile response functions evaluated for the respective AoAs and AoDs at the IRS as explained in the following. 

\begin{figure}\vspace{-0.5cm}
	\centering
	\includegraphics[width=0.5\textwidth]{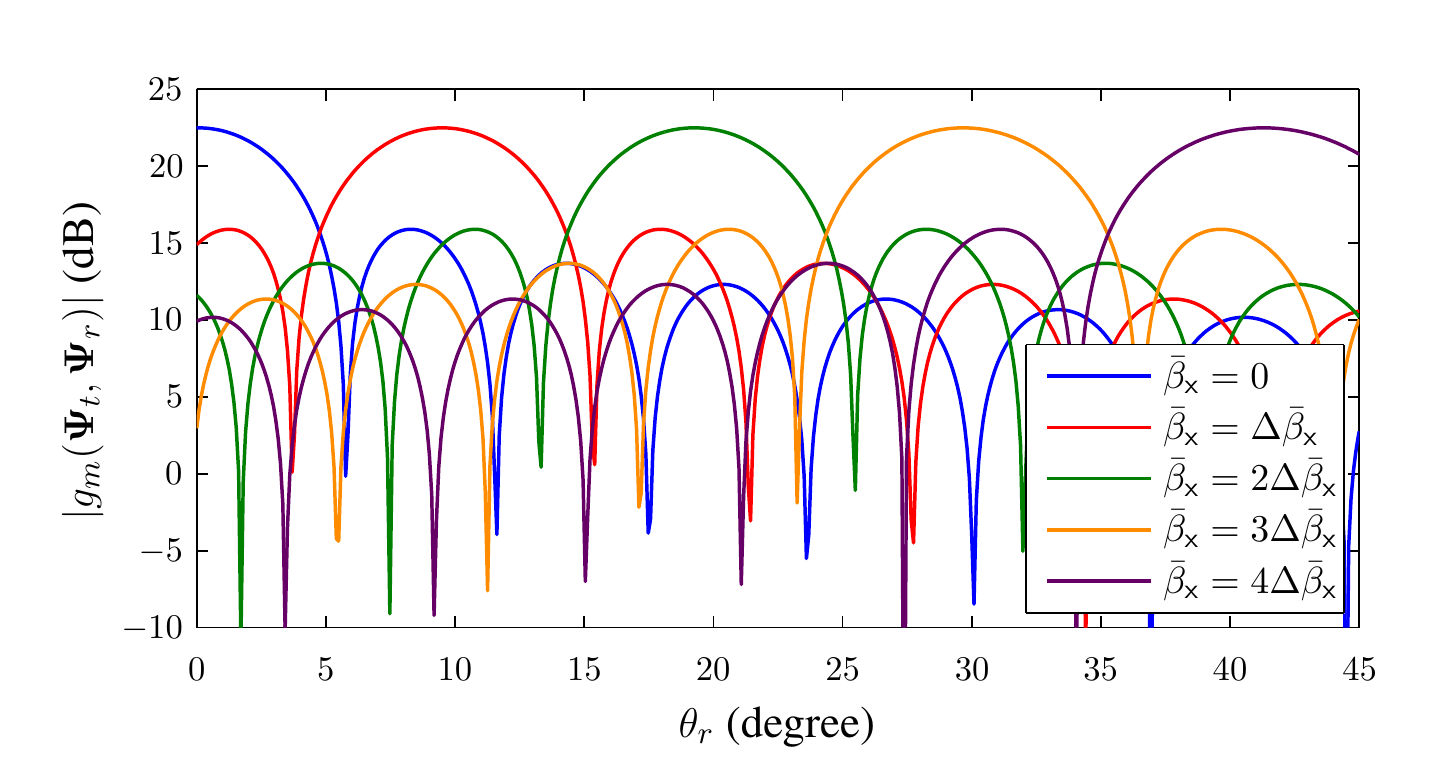}
	\vspace{-0.7cm}
	\caption{$|g_m(\boldsymbol{\Psi}_t,\boldsymbol{\Psi}_r)|$ in dB vs. $\theta_r$  for $(\theta_t,\phi_t,\varphi_t)=(0,0,0)$,  $\phi_r=0$, $L_\x=L_\y=10\lambda$, $d_\x=d_\y=L_e=\frac{\lambda}{2}$,  $\rho_{\rm eff}=0.5$,  $\bar{\beta}_\y=\bar{\beta}_0=0$,  and $\Delta \bar{\beta}_\x=\frac{\sqrt{2}}{16}$. \vspace{-0.5cm}} 
	\label{Fig:Codebook}
\end{figure}

We assume that the IRS can select for each tile one of the possible transmission modes corresponding to one phase-shift function $\beta(x,y)$ in Proposition~\ref{Prop:Continuous}.  A finite number of transmission modes, $M$, can be designed to realize reflection along different directions. 
 Let $s_{n,m}\in\{0,1\}$ denote a binary variable which is equal to one if the $m$-th transmission mode  is selected for the $n$-th tile; otherwise, it is equal to zero. Since, at any given time, the IRS can select only one transmission mode for each tile, $\sum_{m=1}^{M}  s_{n,m}=1,\,\,\forall n$, has to hold. Then, matrix $\mathbf{G}^{(j,i)}$ can be expressed as follows
\begin{IEEEeqnarray}{rll}
\mathbf{G}^{(j,i)} = \sum_{n=1}^N  \sum_{m=1}^{M}  s_{n,m} \mathbf{G}_{n,m}^{(j,i)}  
\end{IEEEeqnarray}
with $\big[ \mathbf{G}_{n,m}^{(j,i)} \big]_{n_r,n_t}	 =  g_{n,m}(\boldsymbol{\Psi}_t^{(n_t)},\boldsymbol{\Psi}_r^{(n_r)})$ being the response function of the $n$-th tile for the $m$-th transmission mode evaluated at the $n_t$-th AoA, specified by angle $\boldsymbol{\Psi}_t^{(n_t)}$, and the $n_r$-th AoD, specified by angle $\boldsymbol{\Psi}_r^{(n_r)}$. Assuming that the center of the $n$-th tile is placed at point $(x,y)=(K_\x^{(n)}L_\x,K_\y^{(n)}L_\y)$, where $K_\x^{(n)}$ and $K_\x^{(n)}$ are integer numbers,  $g_{n,m}(\boldsymbol{\Psi}_t,\boldsymbol{\Psi}_r)$ for AoA $\boldsymbol{\Psi}_t$ and AoD $\boldsymbol{\Psi}_r$ can be expressed as follows
\begin{IEEEeqnarray}{rll}\label{Eq:TilePhaseDiff}
g_{n,m}(\boldsymbol{\Psi}_t,\boldsymbol{\Psi}_r) \!=\! {e}^{{}^{jk[K_\x^{(n)}L_\x A_\x(\boldsymbol{\Psi}_t,\boldsymbol{\Psi}_r) + K_\y^{(n)}L_\y A_\y(\boldsymbol{\Psi}_t,\boldsymbol{\Psi}_r)]}} \!\!\!\!\! g_{m}(\boldsymbol{\Psi}_t,\boldsymbol{\Psi}_r), \nonumber
\end{IEEEeqnarray}
where $g_{m}(\boldsymbol{\Psi}_t,\boldsymbol{\Psi}_r)$ is the response function of the reference tile centered at the origin with $K_\x^{(n)}=K_\y^{(n)}=0$ in transmission mode $m$, which is given in Proposition~\ref{Prop:Continuous}. Each tile can be configured to support reflection along different directions, which is realized by properly choosing the phase shift function $\beta(x,y)$ in \eqref{Eq:BetaContinuous}. In particular, rewriting  \eqref{Eq:BetaContinuous} as $\beta(x)=2k\bar{\beta}_\x x+\bar{\beta}_0/(2\pi)$ and  $\beta(y)=2k\bar{\beta}_\y y+\bar{\beta}_0/(2\pi)$, different transmission modes can be realized by assuming different values for  normalized parameters $\bar{\beta}_i\in[-1,1],\,i\in\{0,\x,\y\}$. For instance, Fig.~\ref{Fig:Codebook} shows the amplitude of the tile response function $|g_m(\boldsymbol{\Psi}_t,\boldsymbol{\Psi}_r)|$ for different transmission  modes $m$ generated by $\bar{\beta}_\x=0,\frac{\sqrt{2}}{16},\frac{\sqrt{2}}{8},\frac{3\sqrt{2}}{16},\frac{\sqrt{2}}{4}$ and $\bar{\beta}_\y=\bar{\beta}_0=0$ when the incident wave is normal to the surface  and the observation point lies in the $x-z$ plane. For the considered example, the transmission modes cover a range of $[0,\pi/4]$ for the elevation angle of the reflected EM wave.

\begin{figure}[t] 
	\centering
	\includegraphics[width=0.4\textwidth]{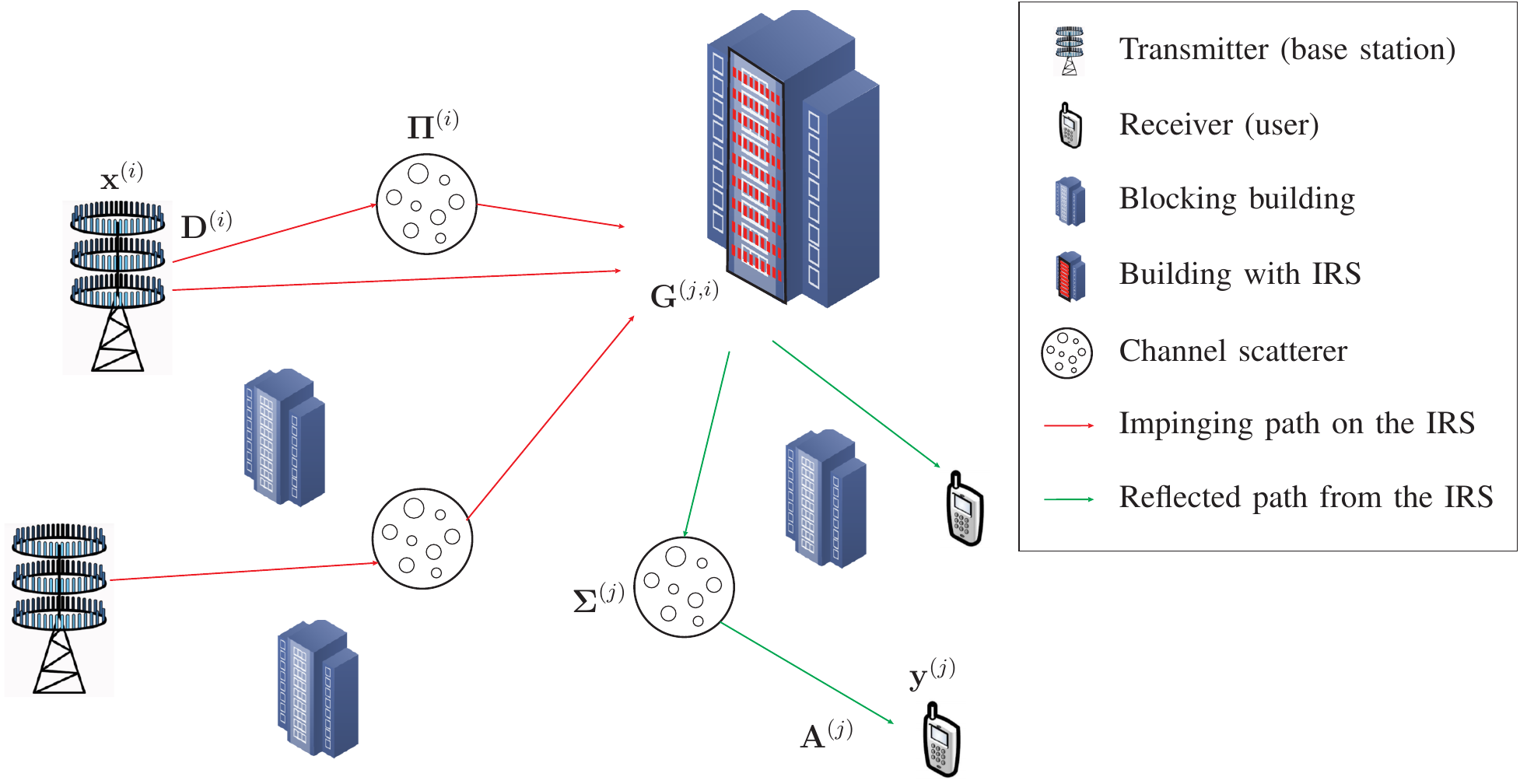}\vspace{-0.2cm}
	\caption{Schematic illustration of an IRS-assisted wireless system. Only the IRS-guided paths are shown for clarity of presentation. \vspace{-0.3cm}}
	\label{Fig:SysMod}
\end{figure}

\section{Conclusions and Outlook}

In this paper, we developed a physics-based channel model for IRS-assisted wireless systems by partitioning the $Q$ IRS unit cells into $N$  tiles and derived the tile response function by solving the corresponding integral equations for electric and magnetic vector fields assuming given unit cell phase shifts at each tile. The proposed model allows the design of scalable optimization algorithms for the joint design of the tile transmission modes and the other system parameters. For example, in the journal version of this paper \cite{najafi2020intelligent}, we jointly optimize the beamforming matrix of the base station (BS) and the tile transmission modes of an IRS-assisted multiuser downlink system. 
Thereby, complexity does not scale with the number of unit cells, $Q$, and the number of possible phases per unit cell, but with the number of tiles, $N$, and the number of transmission modes, $M$.
Assuming the BS employs a large IRS with $3600$ sub-wavelength unit cells to serve two users, we show in \cite{najafi2020intelligent} that the performance does not significantly improve if the number of tiles is increased beyond $N=9$. Thus, the resulting complexity is much lower compared to the conventional approach where the phases of the $Q$ unit cells are directly optimized. This result motivates the application of the proposed IRS-assisted channel model for the design of scalable algorithms for different wireless network architectures and different design goals in future research.


\bibliographystyle{IEEEtran}
\bibliography{References}
\end{document}